\documentclass[preprint2]{aastex}
\usepackage{multirow}
\usepackage{mathptmx}

\begin{document}

\def\dbpsr{PSR\,J0737$-$3039}
\def\dbpsra{PSR\,J0737$-$3039A}
\def\dbpsrb{PSR\,J0737$-$3039B}
\def\psra{PSR\,A}
\def\psrb{PSR\,B}
\def\mos {\emph{EPIC-MOS}}
\def\pn {\emph{EPIC-pn}}
\def\xmm {\emph{XMM-Newton}}
\def\cha {\emph{Chandra}}
\def\flux {\mbox{erg cm$^{-2}$ s$^{-1}$}}
\def\lum {\mbox{erg s$^{-1}$}}
\def\msun{$M_{\odot}$}
\def\farcm{\hbox{$.\mkern-4mu^\prime$}}
\def\farcs{\hbox{$.\!\!^{\prime\prime}$}}
\def\fsecs{\hbox{$.\mkern-4mu^{s}$}}
\def\deg{$^{\circ}$}
\def\degc{\hbox{$.\mkern-4mu^{\circ}$}}
\def\gsim{\hbox{\raise0.5ex\hbox{$>\lower1.06ex\hbox{$\kern-0.94em{\sim}$}$}}}
\def\lsim{\hbox{\raise0.5ex\hbox{$<\lower1.06ex\hbox{$\kern-0.94em{\sim}$}$}}}

\newcommand{\Bf}{{magnetic field}}
\newcommand{\NS}{neutron star}
\newcommand{\ms}{magnetosphere}

\newcommand{\be}{\begin{equation}}
\newcommand{\ee}{\end{equation}}
\newcommand{\nn}{\mbox{} \nonumber \\ \mbox{} }
\newcommand{\ba}{\begin{eqnarray}}
\newcommand{\ea}{\end{eqnarray}}

\title{Long Term Study of the Double Pulsar J0737-3039 with \xmm: pulsar timing}

\shorttitle{}
 \shortauthors{}

\author{M. N. Iacolina\altaffilmark{1}, A. Pellizzoni\altaffilmark{1}, E. Egron\altaffilmark{1}, A. Possenti\altaffilmark{1}, R. Breton\altaffilmark{2}, M. Lyutikov\altaffilmark{3}, M. Kramer\altaffilmark{2,}\altaffilmark{4}, M. Burgay\altaffilmark{1}, S. E. Motta\altaffilmark{5}, A. De Luca\altaffilmark{6}, A. Tiengo\altaffilmark{6,}\altaffilmark{7,}\altaffilmark{8}}
\email{iacolina@oa-cagliari.inaf.it}
\altaffiltext{1}{INAF~-~Osservatorio Astronomico di Cagliari, Via della Scienza 5, I-09047 - Selargius (CA), Italy}
\altaffiltext{2}{Jodrell Bank Centre for Astrophysics, The University of Manchester, Manchester, M13 9PL, UK}
\altaffiltext{3}{Department of Physics, Purdue University, 525 Northwestern Avenue, West Lafayette, IN 47907-2036, USA}
\altaffiltext{4}{Max-Planck-Institut f\"{u}r Radioastronomie, Auf dem H\"{u}gel 69, 53121 Bonn, Germany}
\altaffiltext{5}{University of Oxford - Department of Physics, Astrophysics - Keble Road, Oxford OX1 3RH, UK}
\altaffiltext{6}{INAF~-~Istituto di Astrofisica Spaziale e Fisica Cosmica Milano, Via E. Bassini 15, I-20133 Milano, Italy}
\altaffiltext{7}{Istituto Universitario di Studi Superiori, Piazza della Vittoria 15, I-27100, Pavia, Italy}
\altaffiltext{8}{INFN, Sezione di Pavia, via A. Bassi 6, I-27100, Pavia, Italy}

\begin{abstract}
The relativistic double neutron star binary \dbpsr\ shows clear evidence of orbital phase-dependent wind-companion interaction,  both in radio and X-rays. 
In this paper we present the results of timing analysis  of  \dbpsr\  performed during  2006 and 2011 \xmm\ Large Programs that collected $\sim$20,000 X-ray counts from the system.
We detected pulsations from \dbpsra\ (\psra) through the most accurate timing measurement obtained by \xmm\ so far, the spin period error being of 2$\times$$10^{-13}$ s.
\psra's  pulse profile in X-rays is very stable  despite  significant relativistic spin precession  that occurred within the  time span of observations. This yields a constraint on the misalignment between the spin axis and the orbital momentum axis $\delta_{\rm A} \approx6.6^{+1.3}_{-5.4}$ deg, consistent with estimates based on radio data.
We confirmed pulsed emission from \dbpsrb\ (\psrb) in X-rays even after its disappearance in radio.
The  unusual phenomenology of \psrb's X-ray emission includes orbital pulsed flux and profile variations as well as a loss of pulsar phase coherence on time scales of years.
We hypothesize that this is due to the interaction of \psra's wind with \psrb's magnetosphere and orbital-dependent penetration of the wind plasma onto  \psrb\ closed field lines.
Finally, the analysis of the full \xmm\ dataset provided evidences of orbital flux variability ($\sim$7\%) for the first time, involving a bow-shock scenario between \psra's wind and \psrb's magnetosphere.
\end{abstract}


\section{Introduction}

 \dbpsr\  \citep{bdp+03,lbk+04} is still the most interesting binary system allowing studies on plasma physics, general relativity and matter in the strong-field regime \citep{ksm+06,kw09}.
This is the only known double neutron star system in which both neutron stars have been detected as pulsars. Its orbital period is 2.4 hr, and it contains a mildly recycled pulsar with a spin period of 22.7 ms (\dbpsra\ - hereafter \psra), and a 2.77 s pulsar (\dbpsrb\ - hereafter \psrb).

The uniqueness of this system derives from its highly relativistic nature and from the presence of the two interacting pulsars separated by only $\sim$3 light-seconds whose orbital plane is observed nearly edge-on ($i \approx 88.7^{\circ}$; \citealt{rkr+04}). This has allowed astronomers and theoretical physicists to undertake different kinds of studies never carried out before and obtain new proved results on many lines of researches.
The works of \citet{bdp+03} and \citet{kkl+04} led to an upward estimate of the coalescence rate of \NS\ binaries.
The angle between the \psra's spin axis and the orbital angular momentum was constrained to a small value ($\delta \lsim 4.7^{\circ}$) by \citet{mkp+05} and \citet{fsk+13}. 
Further, while \citet{mll+04c} discovered a strong modulation of pulsed flux density of \psra\ with the periodicity of \psrb\ during the \psra's radio eclipse (when it passes behind \psrb), the studies of \citet{bkk+08} allowed to measure \psrb's spin-orbit precession.
\citet{pmk+10} presented the evolution of \psrb's radio emission, providing an update to \citet{bpm+05} by describing the changes in the pulse profile and flux density over five years of observations, culminating in \psrb's  radio disappearance in March 2008. Over this time, the pulse profile evolved from a single to a double peak, most likely due to relativistic spin precession coupled with an elliptical beam.
In this context, \psrb's radio reappearance is expected to happen at the earliest in $\sim$2035 with the same part of the beam - if we assume a symmetric bipolar beam shape we should have detected \psrb\ in $\sim$2014 -  \citep{pmk+10,ll14}. However, the observed decrease in flux over time and \psrb's disappearance is difficult to explain by current models and may be due to the changing influence of \psra\ on \psrb.
For a comprehensive review of the most interesting scientific results obtained from the study of the Double Pulsar in the radio band so far, see e.g. \citet{ks08} and \citet{kw09}.

High-energy (X-ray and $\gamma$-ray) observations provide a fundamental complement to radio studies for what concerns the physics of the magnetospheric and surface emissions and mutual interactions in the close environment of the two neutron stars \citep{mcb+04a, pda+04, cpb04, cgm+07, ptd+08, prm+08, gkj+13}. High-energy studies of \dbpsr\ could for example reveal an intra-binary shock caused by the relativistic particles wind coming from \psra\ and impacting on the magnetosphere of \psrb\ \citep{liu04}. Furthermore, it is possible to obtain information on the peculiar magnetospheric emission mechanisms of the two close orbiting {\NS}s. In particular, detection and monitoring of \psrb\ in X-rays is of paramount importance after its disappearance in the radio band, providing a unique way to keep monitoring and roughly timing this neutron star.

\citet{ptd+08} reported the results of a $\sim$230 ks long X-ray observation of \dbpsr\ obtained  within the framework of \xmm\ "Large Programs" in October 2006. The analysis of $\sim$5,600 source photons obtained from this observation confirmed the detection of pulsed  X-ray emission from \psra\ (originally revealed by \cha; \citealt{cgm+07}).
\psra\ shows peculiar properties with respect to other known recycled pulsars (e.g. the softest non-thermal spectrum) possibly due to the particular evolutionary history of a double neutron star system. 

For the first time, X-ray pulsed emission from \psrb\ was also detected with good confidence ($\sim$200-300 pulsed counts) around the ascending node of the orbit, though poorly constrained in terms of flux, light curve shape and spectrum. Due to its low rotational energy loss (\psrb\ spin down energy is $\dot{E}_{\rm \psrb} \simeq 1.7 \times 10^{30}$ erg sec$^{-1}$), X-ray emission from \psrb\ can be only powered by an external source, i.e.\,\,the spin-down energy from \psra\ ($\dot{E}_{\rm \psra} \simeq 6 \times 10^{33}$ erg sec$^{-1}$) impacting \psrb's magnetosphere. In this scenario, it was estimated that $\sim$2\% of the rotational energy loss from \psra\ is converted in (possibly thermal) radiation from \psrb. The actual energy transfer and emission mechanisms are still matter of debate \citep{zl04, liu05, lt05, bkm+12}.

\citet{ptd+08} found no signs (e.g. orbital flux modulation and/or a significant non-thermal component in \psra's off-pulse) of the presence of X-ray emission from a bow-shock between \psra's wind and \psrb's magnetosphere invoked in various models  \citep[e.g.][]{liu04,absk05} to explain the occultation of radio emission of \psra\ at inferior conjunction of \psrb. The upper limit on the flux of such a shock component constrains the wind magnetization parameter $\sigma_{\rm M}$ of \psra\ to values $>$1, much higher than that predicted by the radio occultation of \psra\ due to the presence of a magnetosheath (the shocked layer of \psra's wind at \psrb's magnetosphere). 
The absorption causing \psra's occultation occurs within \psrb's magnetosphere which retains enough plasma to produce an eclipse \citep{lt05, bkk+08, bkm+12}.

In order to shed light and deepen the above scenarios, and monitor the system evolution in X-ray on a time scale of years, a new $\sim$360 ks \xmm\ observation was performed in October 2011. In this paper we present the overall timing analysis results of the full $\sim$590~ks \xmm\ dataset on \dbpsr\ including both 2011 and 2006 observations. Spectral analysis results are instead reported by Egron et al (in prep.).

\section{Observations and data reduction}

\begin{deluxetable}{lllccc}
\tabletypesize{\scriptsize}
\tablecolumns{1}
\tablewidth{0pt}
\tablecaption{\label{table:2011obs} \xmm\ observations of \dbpsr\ for the 2011 Large Program$^{\rm d}$}
\tablehead{
\colhead{Orbit} & \colhead{Start} & \colhead{{\it EPIC}}  & \colhead{Total} & \colhead{Background}&\colhead{\phantom{1}Exposure$^{\rm c}$}\\
\colhead{ID} & \colhead{Date} & \colhead{Detector}  & \colhead{source} & \colhead{fraction$^{\rm b}$}& \colhead{\phantom{1}}\\
\colhead{} &  &   &  counts$^{\rm a}$& \colhead{(\%)}& \colhead{\phantom{1}(ks)}
}
\startdata
&&&&\\
      &            &{\it pn}   & 3165 & 31.2 & 88\\
2174   & 2011/10/22 &{\it MOS1} & 582 & 10.0 & 104\\
      &            &{\it MOS2} & 686 & 14.4 & 122\\
\\
      &            &{\it pn}   & 2987 & 40.1 & 65\\
2175   & 2011/10/24 &{\it MOS1} & 724 & 22.5 & 84 \\
      &            &{\it MOS2} & 713 & 22.1 & 84 \\
\\
      &            &{\it pn}   & 2603 & 21.3 & 83\\
2176   & 2011/10/26 &{\it MOS1} & 746 & 11.6 & 112 \\
      &            &{\it MOS2} & 733 & 17.0 & 112 \\
\enddata
\tablenotetext{a}{Total counts in the source extraction region.}
\tablenotetext{b}{Percentage of the background on the on-source total counts.}
\tablenotetext{c}{Total good time after dead-time correction and soft proton flare screening.}
\tablenotetext{d}{For the 2006 observation see Table 1 in \citet{ptd+08}.}
\end{deluxetable}

Our work initially focused on the most recent \xmm\ observation of \dbpsr\
performed in October 2011. The Double Pulsar was observed for three consecutive \xmm\ orbits, whose 
lengths and start times are indicated in Table \ref{table:2011obs}, where 
details of observations are summarized. 
The total time span of the observation covered $\sim$41 
binary orbits (363.7 ks). Data were obtained with the European Photon Imaging Camera (EPIC): 
two \mos\ detectors (each made up of an array of 7 CCDs; \citealt{trp+01}) 
and one \pn\ camera (an array of 12 CCDs; \citealt{sbd+01}), operating in 
Small-Window mode (with thin optical filter), 2'$\times$2' field of view
in the central CCD, with a time resolution of 0.3 s for the \mos\ cameras,
and 4'$\times$4' field of view with a 5.67 ms time resolution for the \pn\ 
camera\footnote{Note that \pn\ higher time resolution modes ($\leq 30\  \mu$s) are only 
suitable for very bright sources: 
\protect{\url{http://xmm.esac.esa.int/external/xmm\_user\_support/documenta- tion/uhb\_2.1/XMM\_UHB.html}}}.

Data analysis has been performed using the \xmm\ Science Analysis Software 
(SAS), version 11.0.0. The first step was to process data with {\tt emproc} 
and {\tt epproc}, standard pipeline tasks for \mos\ and \pn\ observations 
respectively. 
In order to reduce the background contamination we selected data with photon pattern 
0-12 for the \mos\ while, for the \pn, pattern 0-4 for energies higher than 
0.4 keV and 0 for energies lower than 0.4 keV.
Part of the data was affected by background flares. We excluded such
episodes by extracting a 10 s time binned light curve in the 0.15-10 keV 
energy range from the whole field of view for each camera, omitting the 
source region, and selecting events below a threshold of 5~$\sigma$ of the 
quiescent rate for the \pn\ and 3~$\sigma$ for the \mos. 

The radius of source extraction region was set to 18'' for the \pn\ and 15'' 
for the \mos\ data.
After testing a range of thresholds for the background, different photon
pattern selections, and source extraction radii, the adopted background
flares rejection and extraction criteria described above
proved to maximize detection significance of the light curves of \psra\ and \psrb\
in the 0.15-10 keV energy range (see \S \ref{sec:psra} and \S \ref{sec:psrb}). 

The resulting good exposure times and counts statistics for the source and the background 
related to the 2011 observations are listed in Table \ref{table:2011obs}.

Data related to the October 2006 observation and published by \citet{ptd+08} were
re-processed using the same method and tasks described above for the new 2011's dataset
in order to consistently and equally process the whole \xmm\ dataset,
before proceeding with the comprehensive long-term timing analysis of \dbpsr.
For details of the 2006 observations ($\sim$230 ks of data) see Table 1 of \citet{ptd+08}.
Note that also in that case both \mos\ detectors and \pn\ camera were operated in 
Small-Window mode.

In order to perform high-precision timing analysis on such a large data span, the filtered events lists
have been solar system barycentred using the recent and accurate JPL DE405 Planetary Ephemeris\footnote{\protect{\url{http://ssd.jpl.nasa.gov/?planet\_eph\_export}}}. Note that this option was not available for the SAS version 7.1.0 at the time of 2006 data reduction (implicitly using obsolete JPL DE200 Ephemeris).

The last step in the data reduction before entering timing analysis was to correct data for the effects of the relativistic orbital motion of the binary system, according to \cite{bt76}, with the same procedure as in \cite{ptd+08}, and using the ephemerides from \citet{ksm+06} for both \psra\ and \psrb.

The \xmm\ Science Operations Centre (SOC) reported that the observation of \dbpsr\ in
revolutions 2174, 2175 and 2176 originally suffered from time stamp drifts at the Perth Ground Station,
claiming possible corruption on the time dependent analysis of the data.
In fact, we performed an early quick-look timing analysis providing a coarse light curve of \psra\
that revealed an evident loss of signal coherence among and within different \xmm\ revolutions.
In January 2012 the \xmm\ SOC was able to fully fix the problem regenerating
the Time Correlation file and re-ingesting the ODFs in the archive.
The timing analysis presented in the following sections refers to the last version
of the ODFs (ID 0670810101/201/301, release 2) in the \xmm\ public archives, apparently free from
any further time stamp problem.

\section{Long-term X-ray timing of \dbpsra}\label{sec:psra}
The combination of 2006 and 2011 datasets provided $\sim$13,500 \pn\footnote{The \mos\ data cannot be used to perform  \psra\ timing because of the relatively poor time resolution of the instruments, over an order of magnitude higher
than the pulsar spin period.} counts from the source region suitable for pulsar timing analysis.

\begin{figure*}[!t]
\centering
\includegraphics[angle=00,height=7.2cm,width=.465\textwidth]{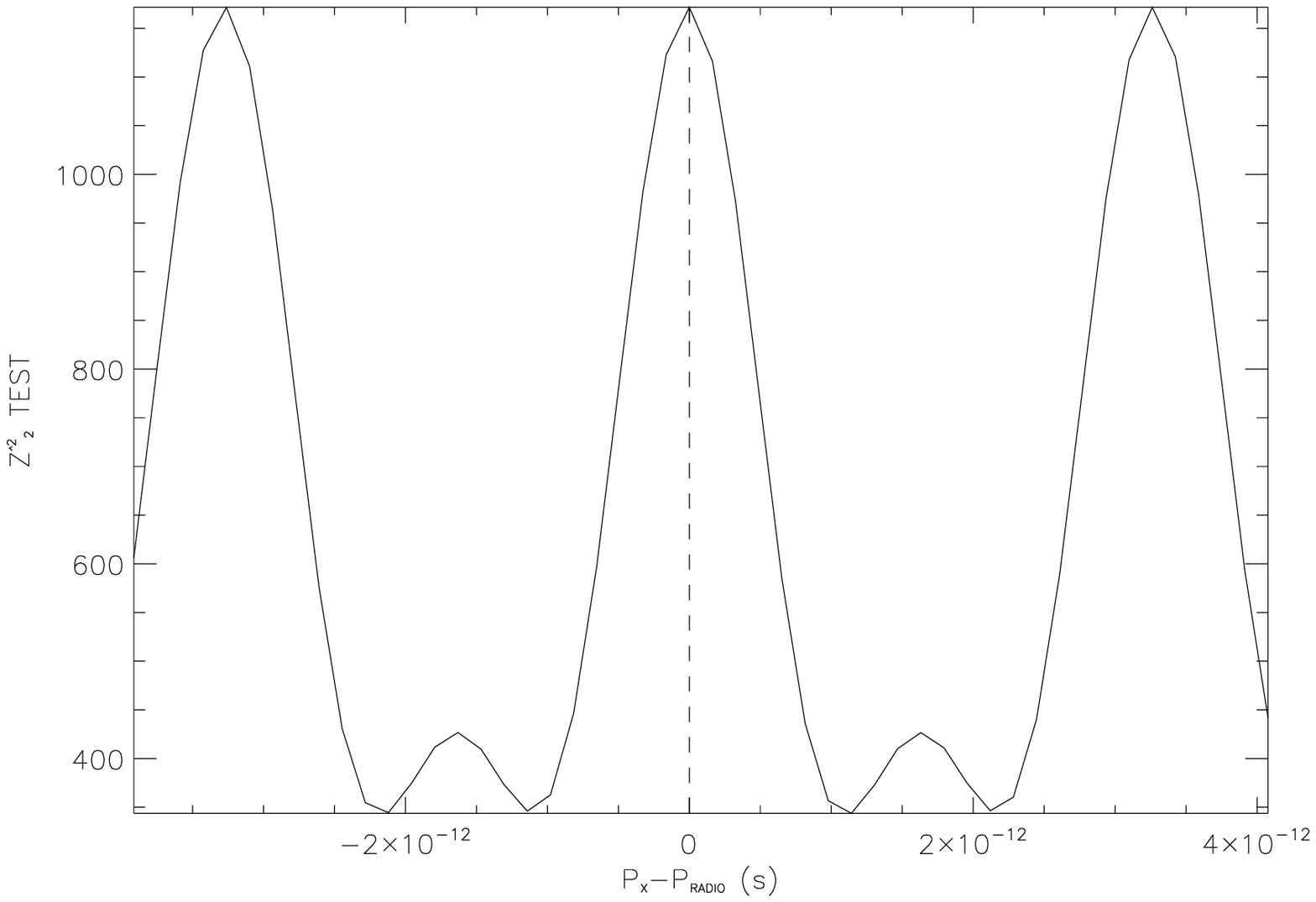}
\includegraphics[height=7cm, width=.53\textwidth,angle=00]{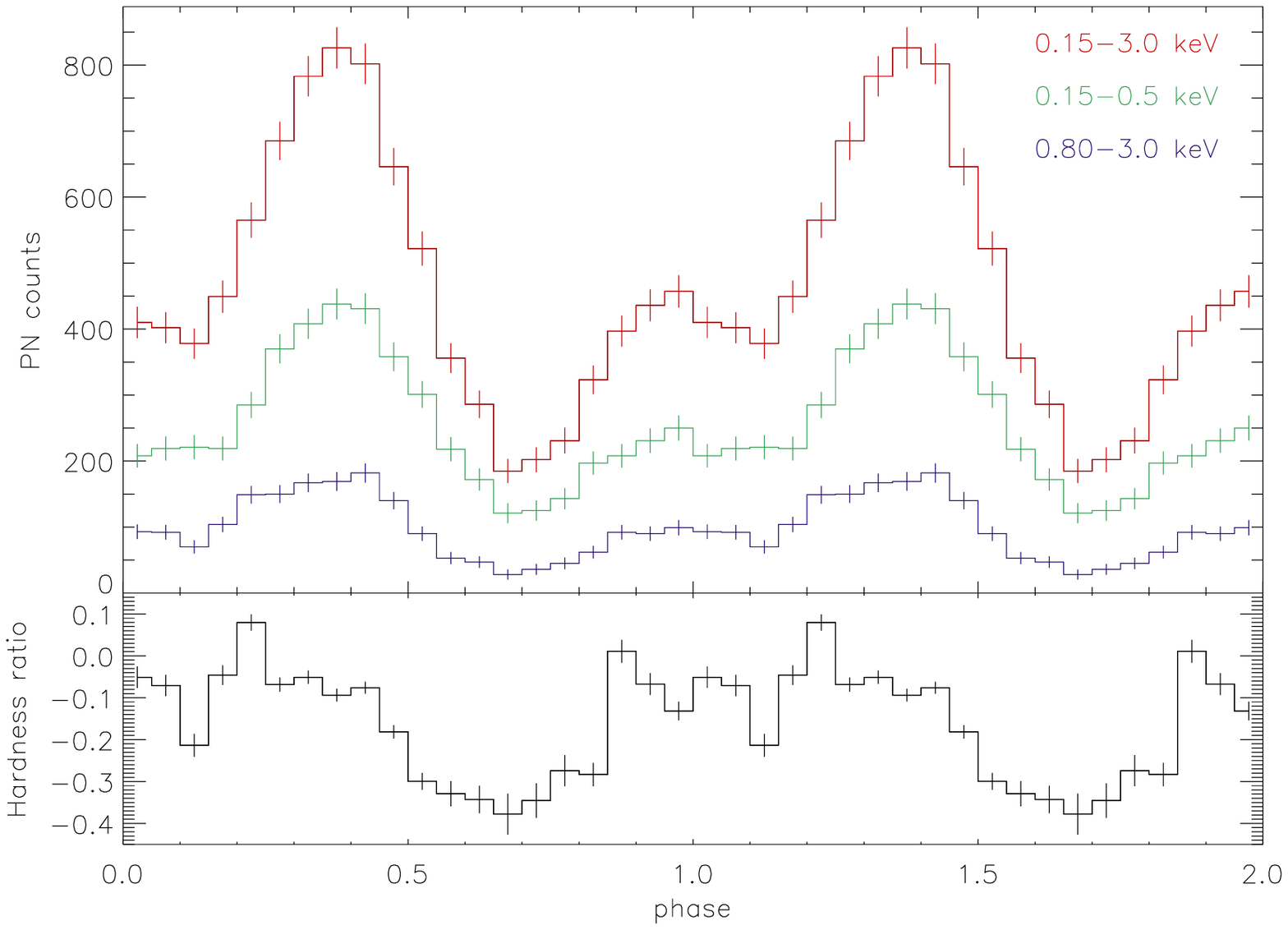}
\caption{{\it Left}: Result of the pulsar spin period search trials around \psra's radio ephemeris solution.
The most significant X-ray pulsed detection corresponds to a $Z^2_2$-test value of $\sim$1200 and is precisely aligned (within $\sim$10$^{-13}$~s) with the radio period indicated by the dashed vertical line. {\it Top-right}: Background subtracted light curve of \psra\ for the energy band 0.15-3.0 keV ({\it in red}) and two sub-bands at low (0.15-0.5 keV, {\it green}) and high (0.8-3.0 keV, {\it blue}) energies, resulting from epoch folding of all available \pn\ data (2006+2011). {\it Bottom-right}: Phase-resolved hardness ratio between the hard (0.8 - 3.0 keV) and soft (0.15 - 0.3 keV) energy bands. }
\label{fig:psra_tot}
\end{figure*}

We searched for pulsar signal in a wide spin period interval centered at the value predicted by the radio ephemeris from \citealt{ksm+06} (keeping all the other ephemeris values fixed) in order to check for possible discrepancies among X-ray and radio timing solutions. We applied the bin independent $Z_{\rm n}^2$ test statistics \citep{bbb+83} to the light curves obtained by folding the data at each trial period.
The adopted period search resolution was $\delta P  = P^2/(T_{\rm obs} n_{\rm bin})=2\times10^{-13}$~s, where $P$ is \psra's spin period, $T_{\rm obs}$ is the time span of the whole observation and $n_{\rm bin}$ the number of bins used to fold the light curve ($n_{\rm bin}$=20 is a suitable choice given the relatively moderate \pn\ time resolution).
The results are shown in Fig.\,\ref{fig:psra_tot} (left panel) for a $Z_{\rm n}^2$ test with $n = 2$ harmonics. The curve presents several peaks corresponding to different trial frequency values. It can be explained considering that the dataset is not continuous, but composed of five \xmm\ orbits related to different observing epochs. In particular, we have a big time gap in the data between the 2006 and 2011 Large Programs and a total of three additional gaps between the two and the three \xmm\ orbits of 2006 and 2011 data respectively. This introduced some periodicities featuring the periodogram as an "interference+diffraction"-like diagram.
The strongest signal, providing $Z_2^2$$\sim$1200 with an almost null probability to be obtained by chance
even accounting for the over $10^4$ period trials, occurred for the period $P_{\rm A,X} = 22.6993785996(2)$~ms referred to the epoch 53156.0 MJD of the radio ephemeris (the error quoted in parenthesis corresponds to the resolution of the search). This value is consistent with $P_{\rm A,radio} = 22.699378599624(1)$~ms, expected at the same epoch. 

It is worth noting that \psra's radio ephemeris from \citet{ksm+06} is still valid several years beyond the range of the timing data (MJDs 52760-53736). This is due to the high-precision measurements of Keplerian and post-Keplerian parameters of the binary orbit, and to the high-stability of the millisecond pulsar not affected by glitches or significant timing noise.

The pulse profile corresponding to the above best timing solution is shown in Fig.\,\ref{fig:psra_tot}, top-right panel. The background-subtracted light curve is shown for three different energy ranges: the band that maximizes the signal to noise ratio (0.15-3.0 keV) and two sub-bands highlighting the behaviour at low (0.15-0.5 keV) and at high-energy (0.8-3.0 keV). Error bars are calculated according to the expression $\sigma_{\rm i}=(C_{\rm i}+B_{\rm i} f^{2})^{1/2}$, where $C_{\rm i}$ and $B_{\rm i}$ are the total counts and the background counts for each bin of the light curve, and $f$ is the ratio between the source and background extraction area ($f\simeq0.1$).
The pulsed counts can be estimated through the expression ${\rm PF}\equiv (C_{\rm tot} - n_{\rm bin} N_{\rm min})$
(and its associated error $\sigma_{\rm PF} = (C_{\rm tot} + n_{\rm bin}^2 N_{\rm min})^{1/2} $), 
where $C_{\rm tot}$ are the total counts and $N_{\rm{min}}$ are the counts from the bin of the minimum of the light curve. 
For the full 0.15-3.0 keV energy range the pulsed flux is $F_{\rm \psra} \approx 5.1^{+1.2}_{-0.9}\times 10^{-14}$ erg cm$^{-2}$ s$^{-1}$ (assuming
the best fit two-component model PL+BB, as in \citealt{ptd+08}), corresponding to a pulsed counts fraction of $60\% \pm 4\%$ ($n_{\rm bin}$ = 20) with respect to the total background-subtracted source counts. 
Assuming a distance $d_{\rm \psra} \approx$ 1.1 kpc \citep{vwc+12}, the luminosity of \psra\ ($L_{\rm \psra} = \Omega_{\rm A} d^2_{\rm \psra} F_{\rm \psra}$) results in $5.9\times 10^{29}\Omega_{\rm A}$~erg~s$^{-1}$, with a X-ray efficiency $L_{\rm \psra}/\dot{E}_{\rm \psra} \sim 10^{-4}\Omega_{\rm A}$, where $\Omega_{\rm A}$ ($\sim$ 1-10 steradians) is the pulsar beam solid angle, a value in agreement with the expectation for the recycled millisecond pulsar if $\Omega_{\rm A}>1$ \citep{bt99,pccm02}. 

The spectral hardness ratio as a function of the pulsar phase was calculated according to the expression $(H_{\rm i} - S_{\rm i}) / (H_{\rm i} + S_{\rm i})$, where $H_{\rm i}$ and $S_{\rm i}$ are the counts in each bin for the hard (0.8 - 3.0 keV) and soft (0.15 - 0.3 keV) energy ranges respectively. The result is shown in Fig.\,\ref{fig:psra_tot}, bottom-right panel, and confirms the behaviour exhibited in \citet{ptd+08}, indicating a softer spectrum in correspondence to the minimum of the pulse profile. To better investigate this feature, we analyzed the pulsed emission in different energy bands. The background-subtracted light curves for four different energy bands (0.15-0.3 keV; 0.3-0.5 keV; 0.5-0.8 keV; 0.8-3.0 keV) show that the pulsed fraction increases from $\sim$50\% to $\sim$70\% as the energy in the intervals increases with a similar trend in 2006 and 2011.

\begin{figure*}[t]
\centering
\resizebox{.6\hsize}{!}{\includegraphics[angle=00]{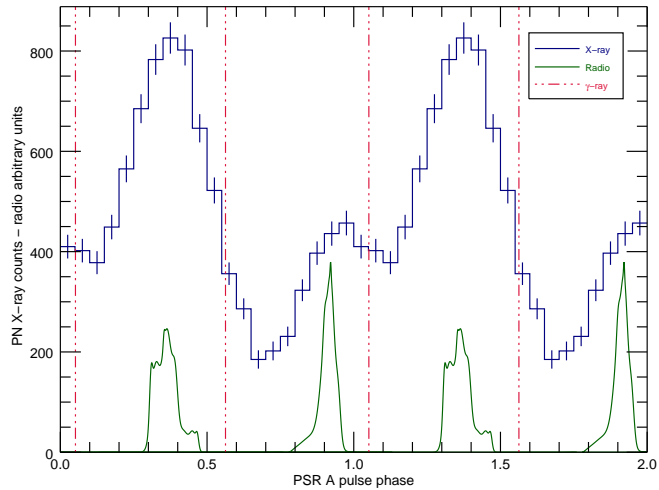}}
\caption{Result of phasing of the \psra's pulsed profile. The X-ray profile obtained for 0.15-3.0 keV is represented by the solid blue line. The green line shows the radio profile at 820 MHz in arbitrary units. While the dotted-dashed red vertical lines indicate the phase of the maximum of the $\gamma$-ray pulses. Two rotational phases are displayed for clarity.}
\label{fig:psraphaGXR}
\end{figure*}

In order to search for a possible temporal evolution of \psra\ pulse profile caused e.g. by relativistic pulsar spin precession (having a period of $\sim$75 years; \citealt{lbk+04}), we compared the light curve obtained in 2006 from the first \xmm\ Large Program \citep{ptd+08} to that obtained from the latest Large Program (2011).  We applied the two-sample Kolmogorov-Smirnov (K-S) test to the time series comparing the whole 2011 and 2006 profiles and the principal and secondary peaks individually, both for the whole energy band (0.15 - 3 keV) and the soft (0.15 - 0.3 keV) and hard (0.8 - 3.0 keV) ranges. Considering all the above mentioned cases, the distribution functions of the two datasets resulted consistent, being the differences significant only at $\sim$2.3~$\sigma$. Moreover, we searched for possible spin phase shift or jumps among (and within) the two datasets (2006 and 2011) in order to check for possible long-term data acquisition system instability or irregular pulsar timing behaviour. 
No significant phase shift was present when planetary ephemeris were coherently used for solar system reference frame conversions. An upper limit of $\sim$0.04 (3~$\sigma$) was computed by artificially introducing 100 phase shifts trial values (step of 0.01) between the two time series before applying the K-S test.
Obtaining instead barycentred times using ephemeris DE200 for the 2006's dataset (the only available in the coeval SAS versions), and then using updated ephemeris DE405 for the new 2011's observation, one introduces an apparent phase shift of $\Delta \phi_{\rm spin}  \simeq$0.07 among the two datasets.

The possible temporal evolution of \psra's pulse profile can also be evaluated, as a cross-check, considering the ratio of the two folded light curves normalized to their respective total counts. The result did not show any modulation, having a $\chi^2_{\rm red}$ significance value of $\sim$2.2~$\sigma$ ($\chi^2_{\rm red}$ applied to the ratio of the time series) and confirming the result obtained from the two-sample K-S test.
Through the same technique we estimated a phase shift between the two light curves of 0.008$_{+0.001}^{-0.006}$ (errors at 1~$\sigma$), slightly improving the upper limit obtained by the K-S test. 

In summary, no significant pulse profile evolution or phase shift can be firmly claimed.

\subsection{Pulse phasing of \dbpsra}\label{sec:psrapha}

In order to better constrain the possible emission mechanisms providing the observed pulsed profile of \psra, it is important to assess the relative multi-wavelength phasing of the pulsed peaks. This will also helps us to further constrain the geometry of the system. Previous works studying the polarization of radio profile \citep{drb+04} showed that a possible interpretation of the double peaked pulsed profile is emission from a single big hollow cone of a nearly aligned rotator. On the other hand, \citet{fsk+13} suggested a double cone emission from an orthogonal rotator. 
In 2007 \citeauthor{cgm+07} compared X-ray and radio pulsed profiles, while \citet{gkj+13} compared radio and $\gamma$-ray ones. In this work we collected the three profiles finding out a X-ray/radio phasing in contrast with the previous results.

Fig.\,\ref{fig:psraphaGXR} shows the X-ray folded light curve in the 0.15-3.0 keV energy band, the 820 MHz radio profile obtained from GBT data at epoch 55855 MJD, and the location of the peak maxima of the $\gamma$-ray profile. 
We obtained that the maximum of the lower radio peak is located at phase 0.36 and the highest X-ray one is at 0.38 with a phase difference of 0.02$\pm$0.03, while the highest radio peak is at 0.92 and the lower X-ray one at 0.98 with a difference of 0.06$\pm$0.03 (the phasing error being determined by the relatively low count statistics of the X-ray light curve).

Our phasing result is shifted of $\sim$0.5 with respect to that reported by \citet{cgm+07} in which the two highest X-ray/radio peaks and the two lower were corresponding.
Phase jumps or shift artifacts in the \xmm\ data are excluded from the clear evidence of peak coherence on the very long data span involving multiple data sets at different epochs. 

Furthermore, we found a separation between the two X-ray peaks of $\sim$0.6$\pm$0.03 ($\sim$216\deg$ \pm$11\deg), while \citet{cgm+07} reported an X-ray peak-to-peak separation of $\sim$182\deg$ \pm$3\deg. In contrast with this latter result, the separation measured in this work is compatible (at least at 2~$\sigma$ level) with the observed separation among the radio peaks ($\sim$200\deg;  \citealt{mkp+05}).

Our analyses suggest that X-ray and radio emission are coming from the same region of the magnetosphere (apparently unrelated to $\gamma$-rays)
while the pattern of the radio emission (single or double cone) cannot be constrained by our result. We note that the absence of bridge emission in both radio and X-ray profiles is compatible with the hypothesis of a single hollow cone, in contrast with \citet{cgm+07} result which would imply a center-filled X-ray cone.


\section{X-ray Emission from \dbpsrb}\label{sec:psrb}

Evidence for pulsations ($\sim$4~$\sigma$) from \psrb\ in the 2006 \pn\ dataset was obtained by  \citet{ptd+08} for a (poorly constrained) orbital phase interval around the ascending node\footnote{i.e. orbital phase interval 0.38-0.63 assuming phase ph$_{\rm A}=0$ as a reference for the passage of \psra\ at the orbit's ascending node} of pulsar's orbit. 

In order to verify and better assess this detection, we first analyzed the new 2011 data alone and then re-analyzed 2006 data as a cross-check of previous results before proceeding with the comprehensive timing analysis of both dataset with a 5-year gap.

\psrb's pulsed emission was investigated using the same period search method adopted for \psra\ and radio ephemeris by \citet{ksm+06}. In this case we could also use the lower time resolution \mos\ data in addition to \pn\ data providing $\sim$20,000 source counts overall (2006+2011 datasets), in principle suitable for timing analysis.

\begin{figure*}[t]
\centering
\resizebox{.48\hsize}{!}{\includegraphics[angle=00]{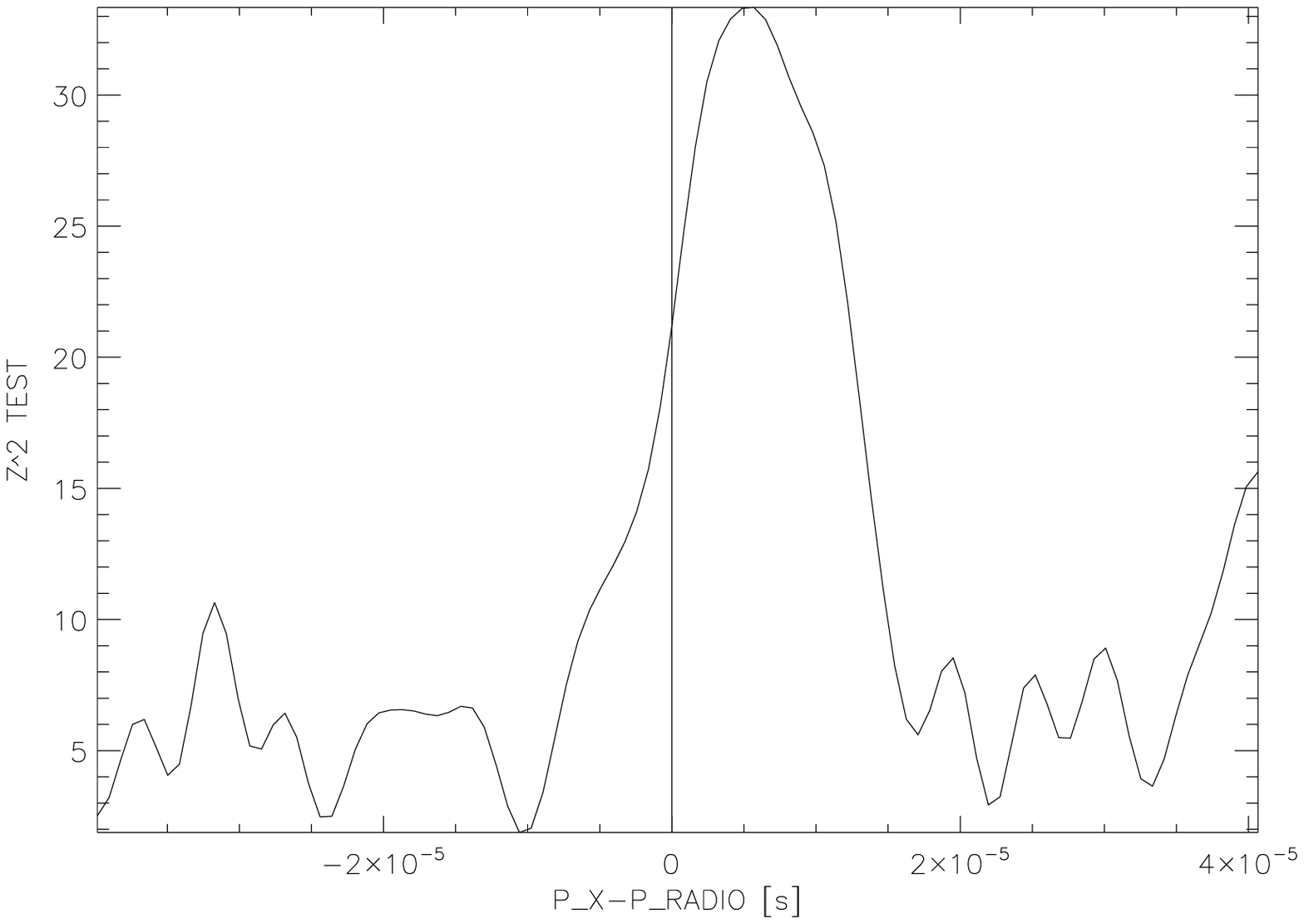}}\hspace{0.1cm}
\resizebox{.48\hsize}{!}{\includegraphics[angle=00]{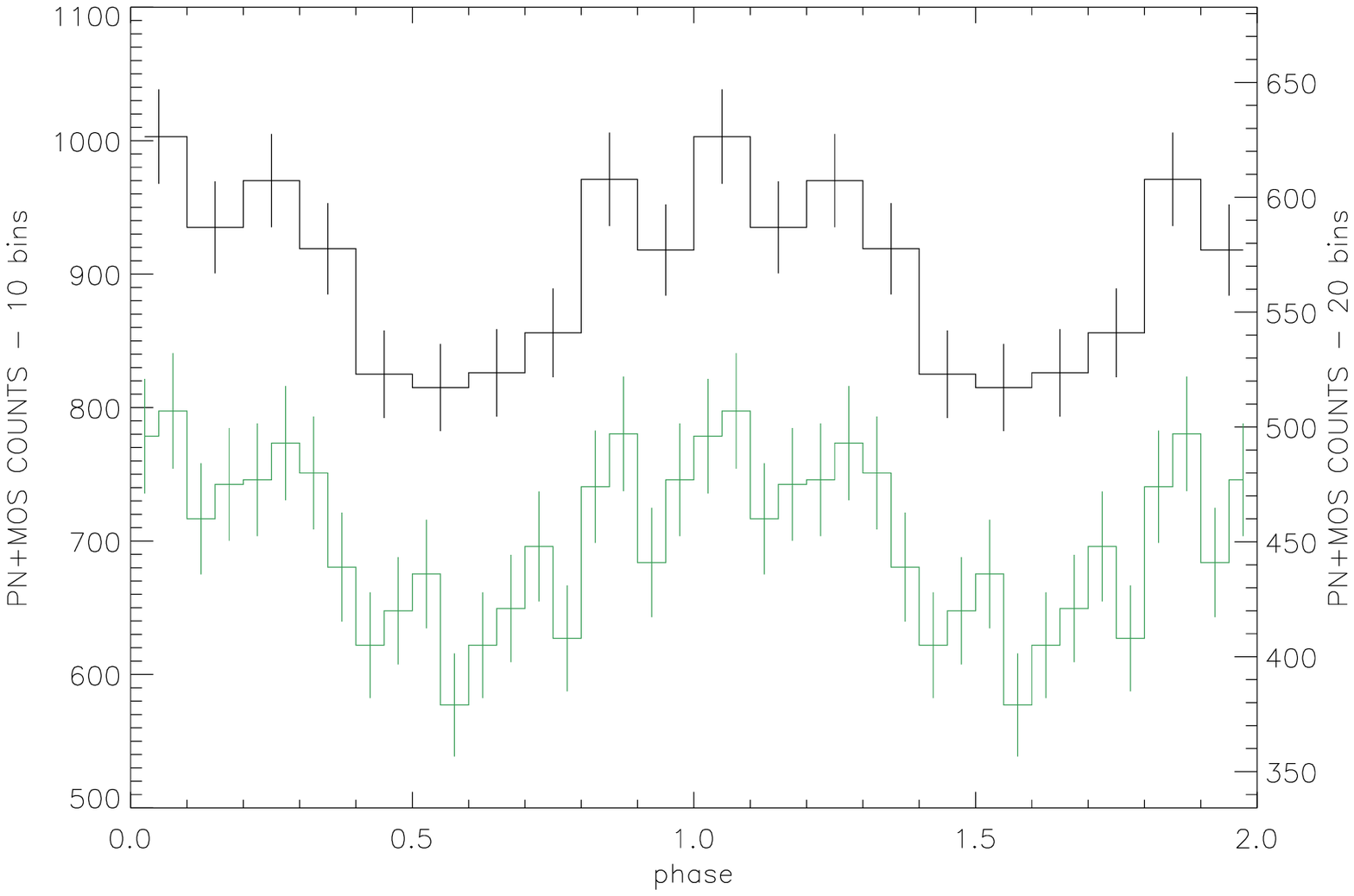}}
\resizebox{.48\hsize}{!}{\includegraphics[angle=00]{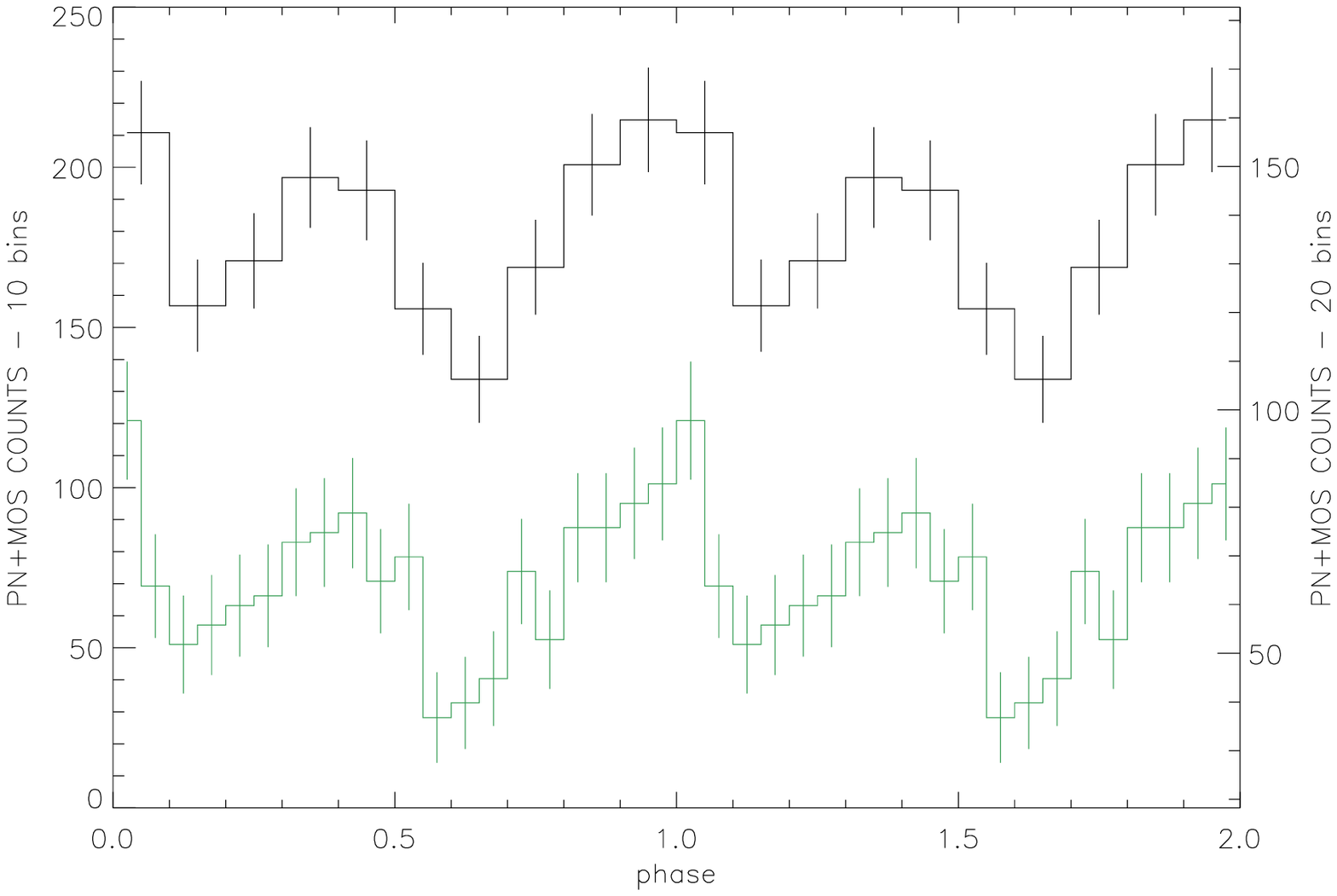}}
\caption{X-ray detection of \psrb. Top left: result of the $Z^2_4$ test distribution for 100 period search trials around \psrb's radio ephemeris. Top right: background-subtracted \psrb\ profile for the 0.15-3.0 keV energy band obtained by folding the 2011 combined \mos\ and \pn\ data (both light curves with 10 bins - {\it in black} - and 20 bins - {\it in green} - are displayed). Bottom: same as top right, but selecting the orbital phase interval $\Delta \phi_{\rm orb} = 0.58 - 0.78$.}
\label{fig:psrb_lc}
\end{figure*}

\begin{figure*}[t]
\centering
\resizebox{.9\hsize}{!}{\includegraphics[angle=00]{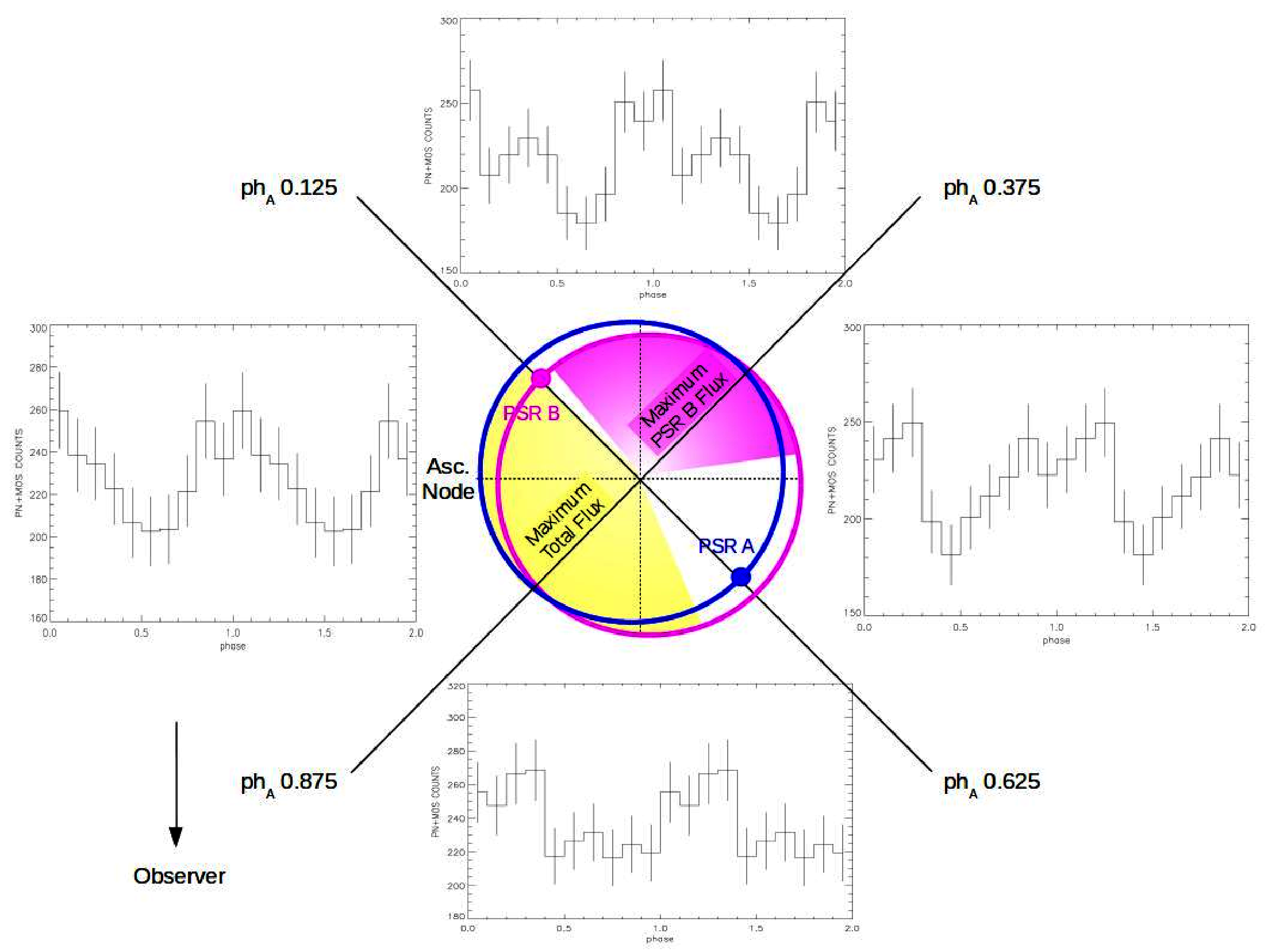}}
\caption{Central panel: scheme of the orbital configuration of \dbpsr\ in 2011 subdivided in four orbital phase intervals containing conjunctions ([0.125,0.375] - [0.625,0.875]) and quadratures ([0.875,0.125] - [0.375,0.625]). The yellow shaded region corresponds to the maximum of the total X-ray flux from the system (see \S \ref{sec:orbital} and Fig.\,\ref{fig:orb_mod}), while the pink area indicates the maximum of the \psrb\ pulsed flux (both shaded regions highlight the phase intervals where \psrb\ is passing). The ph$_{\rm A}$ label in the figure highlights that phase 0 is referred to the \psra\ ascending node. The four plots displayed around the scheme show the \psrb\ (background subtracted) pulse profile obtained by folding the 2011 \mos\ and \pn\ counts belonging to the above mentioned orbital phases for the 0.15-3.0 keV energy band.}
\label{fig:psrb_scheme-lc}
\end{figure*}

The search has been carried out using the $Z^2_{\rm n}$ test (with n ranging from 1 to 5) accounting for detection parameters by \citet{ptd+08}, both for the entire pulsar orbit and also focusing on selected orbital phase intervals (different longitude intervals w.r.t. ascending node). 
The best result of the search was obtained for $n$ = 4 using a period search resolution of  $\delta P = P^2/(n_{\rm bin} T_{\rm obs}$) = 8$\times 10^{-7}$~s ($n_{\rm bin}$=20) for the 2011 data.
The $Z^2_4$ values resulting from 100 period trials around the radio period in the energy band 0.15 - 3.0 keV are plotted in the top left panel of Fig.\,\ref{fig:psrb_lc}. The most significant spin period found in the search is 2.7734665(8) s (referred to the epoch 53156.0 MJD of the radio ephemeris) providing $Z^2_4$ = 33.3 and a $\sim$$4~\sigma$ confidence level (null hypothesis probability of $5\times10^{-5}$). The corresponding background subtracted light curve is shown in the upper right panel of Fig.\,\ref{fig:psrb_lc}. The $\chi^2$ analysis applied to the same timing result indicates a $\chi^2_{\rm red}$ = 2.4 (19 degrees of freedom, $\sim$$3.5~\sigma$). An even better detection significance is obtained applying the $\chi^2$ statistics to a coarser binning (i.e. $\sim$$4.2~\sigma$ detection for $n_{\rm bin}$=10).
The pulsed flux in the 0.15-3.0 keV band is $\sim$1.4$\times$10$^{-14}$ erg cm$^{-2}$ s$^{-1}$, corresponding to a background subtracted pulsed fraction of $\sim$16\%$\pm$5\% (n$_{\rm bin}$ = 20). The related X-ray luminosity is $\sim$1.6$\times$10$^{29}\Omega_{\rm B}$~erg~s$^{-1}$ (where $\Omega_{\rm B}\sim 1$-10 is the pulsar beam solid angle), a value of the same order of the spin-down luminosity of \psrb.

In order to check the robustness of the detection of the relatively weak and peculiar signal from \psrb, we verified that our timing procedure was not affected by systematic errors artificially boosting $\chi^2$ and $Z^2_{\rm n}$ statistics. We replicated the timing procedure using fake radio ephemeris (randomly produced) and we verified that the occurrence of fake detection is compliant with the number of trials. For example, after $\sim$10$^4$ random pulsar search trials, no fake detection $\geq4\sigma$ in both $\chi^2$ and $Z^2$ statistics was found.

Since the light curve of \psra\ shows a larger unpulsed flux fraction in the soft energy band (0.15-0.5 keV), we verified if the detection significance of \psrb\ improves restricting the timing analysis in this band. The result shows a worsening of the signal in the 0.15-0.5 keV band, with a $Z^2_4$ = 14 ($\sim$$1.8~\sigma$). 
On the other hand, in the higher energy band (0.8-3.0 keV) the signal is still significant with a $Z^2_4$ = 24 ($\sim$$3.1~\sigma$).

We also carefully checked how detection significance changes analyzing \pn\ and \mos\ data separately.
Both \pn\ and \mos\ contribute to the detection of the pulsed signal from \psrb\ in the 0.15-3.0 keV band, although the significance of \pn\ light curve is lower than expected ($\chi^2_{\rm red}$ = 1.6) despite it yields about twice as much counts statistics w.r.t. the \mos. This can be explained by the fact that the \pn\ is much more affected by the strong and very soft background emission from \psra. In fact, selecting higher energy ranges (e.g. 0.8-3.0 keV) the significance of \pn\ light curve exceeds that of the \mos\footnote{The $\chi^2$ statistic for $n_{\rm bin}$=10 provides $\chi^2_{\rm red}$ = 2.0 for the \pn,  $\chi^2_{\rm red}$ = 1.8 for the \mos\ and $\chi^2_{\rm red}$ = 3.4 for the combined \pn\ \mos\ light curves.}.

Periodicity was also investigated in four selected orbital intervals around conjunction and quadrature passages of the pulsar\footnote{Orbital phase intervals: {\it quadratures} [-0.125, 0.125] and [0.375, 0.625]; {\it conjunctions} [0.125, 0.375] and [0.625, 0.875] assuming phase ph$_{\rm A}= 0$ as a reference for the passage of \psra\ at the orbit's ascending node and then ph$_{\rm A}$$\simeq 0.5$ for the same passage of \psrb.}. The light curves of \psrb\ obtained by folding \pn\ and \mos\ counts in the selected orbital phase intervals - for the 0.15-3.0 keV energy band - are shown in Fig.\,\ref{fig:psrb_scheme-lc}.
Provided the low pulsed count statistics of the four orbital phase-resolved \psrb's light curves (Z$^2_{\rm n}$ and $\chi^2$ test significance of the pulsed signal $\sim1.8~\sigma$ on average), variations of the pulse profile and pulsed fraction along the orbit cannot be firmly quantified.

After a blind search over a variety of orbital phase intervals (about 20 trials) selected spanning a range of widths and initial phase values, our most significant detection (with null hypothesis probability $\sim10^{-2}$) is obtained in the range 0.58-0.78. This result is shown in the bottom panel of Fig.\,\ref{fig:psrb_lc}. The corresponding two-peaked light curve confirms changes in the pulse profile along the orbit or pulsar phase coherence loss (as also observed in the radio band by \citealt{bdp+03}) maybe due to the peculiar process of illumination of \psrb\ and energy transfer from \psra.

Moreover, the light curve of \psrb\ exhibits pulsed flux variations ($>$50\%) along the orbit peaking in the orbital phase range ph$_{\rm A}$ = 0.65-0.95 (see pink region in Fig.\,\ref{fig:psrb_scheme-lc}). Despite the significance of this result is $< 2 \sigma$ (likely due to the weakness of the signal from \psrb), we think that it is worth to further investigate on this issue.

The measured X-ray period of \psrb\ differs by 5.7$\times 10^{-6}$~s from the expected radio period at the same epoch ($P_{\rm B,radio}$ = 2.77346077007(8)~s). Although the X-ray period search still yields a significant signal ($Z^2_4 \approx$ 22) also at the nominal ephemeris value, the best X-ray pulse period that we find is significantly different from the radio one.
Our high-precision timing results of \psra\ fully exclude instrumental issues or other error sources in our timing procedure.
On the other hand, we have to consider that \psrb's ephemeris errors obtained by \citet{ksm+06} are only valid for the fitted time span of the radio observations due to the fragmented time interval in which \psrb\ is bright (see \citealt{ksm+06} for details). 
Nevertheless,  this discrepancy among X-ray and radio timing of \psrb\ could have a physical origin. Recalling that \psrb\ disappeared in the radio band in 2008 and the radio ephemeris could not be updated since then, we cannot in principle exclude that glitches and/or significant timing noise occurred in the time between the radio and X-ray observations. 
Other viable interpretations should be also considered, e.g. the location of the emitting region far from the pulsar surface can affect barycentric corrections in the binary system. In fact, also the period of \psrb\ detected by \citet{ptd+08} in the 2006 dataset presented a shift of $\sim$$10^{-6}$~s with respect to the radio period.
We verified the above result re-analyzing the 2006 dataset through the new SAS version and updated planetary ephemeris providing comparable results.
We confirmed that in contrast with the 2011 observations, \psrb\ is not detected in the older dataset when considering the whole orbit nor when merging \pn\ and \mos\ data, while the pulsed signal with the same features as in \citet{ptd+08} is found with a comparable significance analyzing \pn\ data in the orbital phase interval 0.38-0.63 (corresponding to the 0.41-0.66 in \citealt{ptd+08}). Pulse profile and spectral variations should then be invoked when comparing \psrb's peculiar behaviour in the two different data spans at the two different epochs.

Finally, we repeated the above analysis for the entire \xmm\ dataset (2006+2011 data). 
Since among 2006 and 2011 observations the relativistic advance of periastron rotated the line of apsides by 84.4\deg, we shifted the binary phases of the second dataset by this value to correct for the effect and effectively sum the counts arising from conjunctions and quadratures. 
Moreover, the relativistic spin precession produces a rotational phase shift which would prevent coherent folding on such a large timescale. We then tried to recover the original phase introducing arbitrary phase shifts in the light curve up to $\phi^{\rm max}_{\rm spin} \approx \Delta T_{\rm X}/P_{\rm prec} \approx 0.07$, where $\Delta T_{\rm X}$ is the time elapsed from the beginning of the 2006 observation and $P_{\rm prec}$ is the relativistic spin precession period (71 years; \citealt{lbk+04}).
The result did not show any modulation of the signal. The light curve evinces an almost flat behaviour at all orbital phases. 

Moreover, merging the two dataset without introducing any phase shift due to the advance of periastron (and then
necessarily neglecting observer-related orbital variability), we also verified that no emission enhancement/decrease deriving from the periastron and apoastron passages is present.

The lack of detection in this search on the overall \xmm\ dataset could possibly be related to the fact that \psrb's ephemeris may not be valid over such a long time span, or to a loss of phase coherence related to \psrb's X-ray emission mechanism connected to the energy transfer from \psra.

\begin{figure*}[t]
\centering
\resizebox{.48\hsize}{!}{\includegraphics[angle=00]{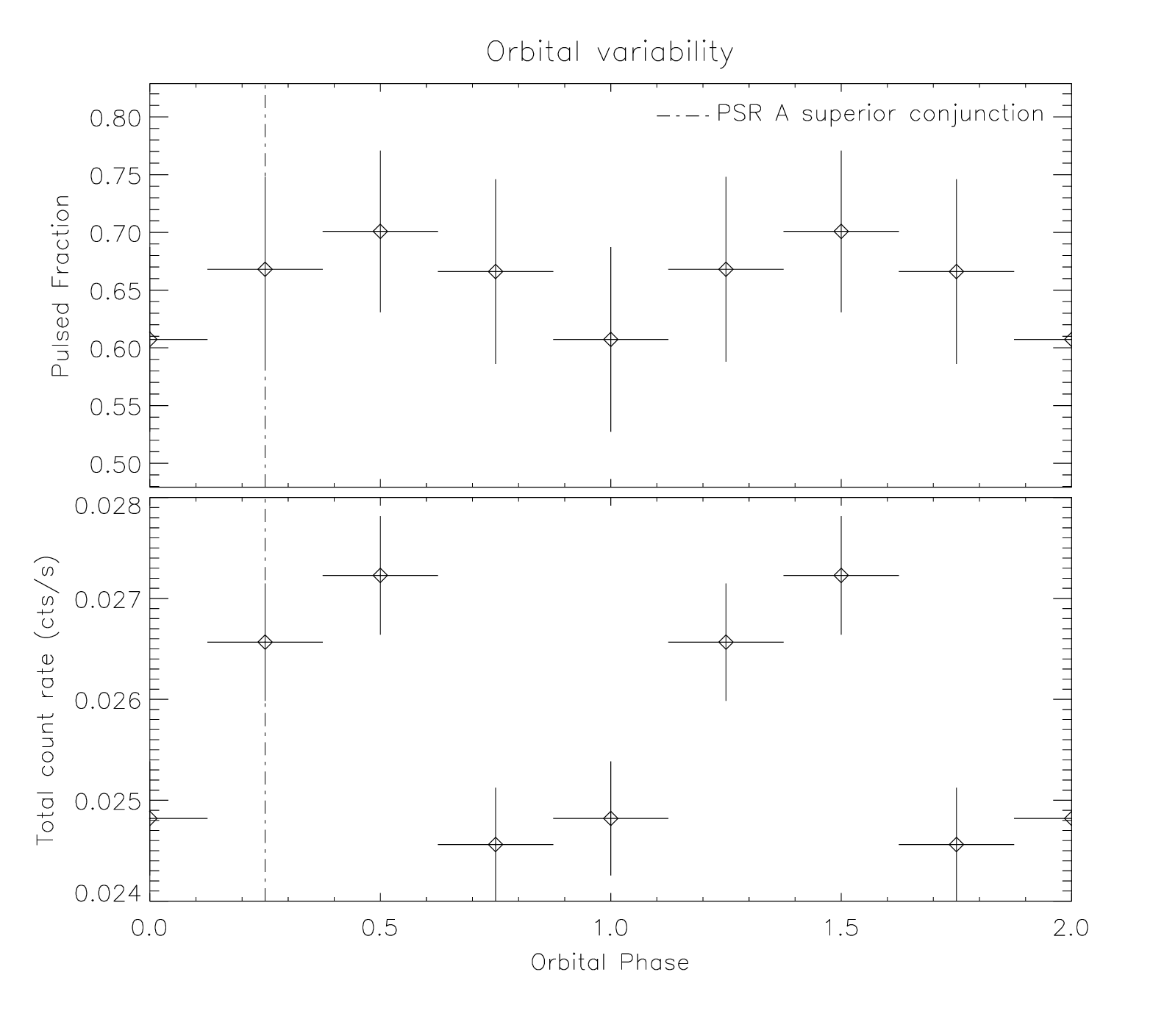}}
\resizebox{.48\hsize}{!}{\includegraphics[angle=00]{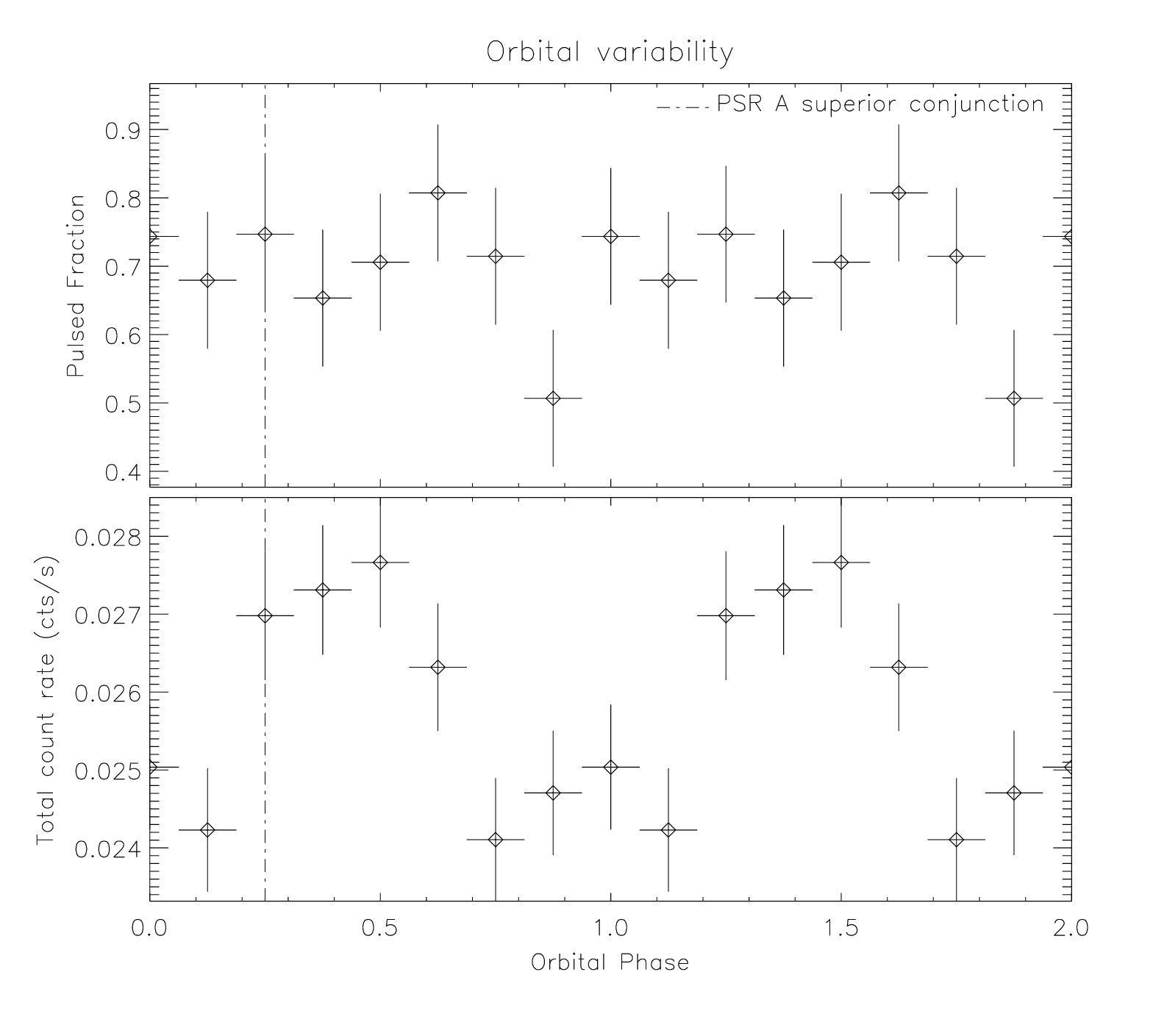}}
\caption{Background-subtracted pulsed fraction of \psra\ ({\it top panels}) and orbital modulation of the total count rate from the system ({\it bottom panels}), in the 0.15 - 3.0 keV energy interval, obtained by folding $\sim$67 orbital periods (2006+2011 \pn\ data) and binning at $\sim$40 min ({\it left}) and $\sim$20 min ({\it right}). The vertical dotted-dashed line represents the superior conjunction of \psra.}
\label{fig:orb_mod}
\end{figure*}

\section{Orbital variability}\label{sec:orbital}

Due to the noticeable evidence of peculiar interactions among the two pulsars both in radio and X-ray, a careful search for orbital flux variability with the strongly improved counts statistics with respect to previous works was a fundamental step in the system's timing analysis.
Even though Pellizzoni et al. (2008) reported a quite stringent upper limit ($<$10\%) on orbital flux variability, the X-ray emission of \psrb\ could possibly be associated with appreciable flux variation as a function of the orbital phase. Moreover, anisotropic high-energy emission could be produced at the interacting layer between \psra's wind and \psrb's magnetosphere (see e.g. Lyutikov (2005), bow-shock model). Figure \ref{fig:orb_mod} bottom panels show the background-subtracted orbital light curves obtained by folding $\sim$67 \dbpsr\ orbits in the energy band 0.15-3.0 keV, together with the corresponding pulsed fraction of \psra\ (2006+2011 \pn\ data) shown in the top panels.
Evidence of moderate orbital modulation of the total count rate is obtained when considering the entire \xmm\ dataset (bottom panels of Fig. \ref{fig:orb_mod}), while no significant detection of variability can be claimed from the analysis of the 2006 or 2011 dataset separately.
The full dataset was processed taking into account the effects of the advance of periastron which occurred between the two observation periods (2006 and 2011) and correcting for their relative orbital phase shift, as we carried out for  \psrb's timing analysis.
The analysis of modulation resulted in a $\chi^2_{\rm red}$ = 3.2, with a null hypothesis probability of 0.3\% ($\sim$3$\sigma$) adopting a $\sim$20 min binning, and $\chi^2_{\rm red}$ = 5.2, with a null hypothesis probability of 0.01\% ($\sim$3.2$\sigma$) for $\sim$40 min binning. The $Z_{\rm n}^2$-test applied to the light curve provided a detection significance of 3.6~$\sigma$ ($Z^2_1$ = 16).
The detection significance slightly improves when considering only the softer energy band (0.15-0.5 keV). The inclusion of \mos\ data in the total flux variability analysis does not change the overall result.

The orbital count rate is clearly higher in the orbital phase range $\sim$0.2-0.6 (phase=0 corresponds to the passage of \psra\ at the ascending node) peaking at phase $\sim$0.5, near the descending node of \psra. The minimum of the orbital emission is at $\sim$0.75, close to \psrb's superior conjunction. We observe approximately the same behaviour in the soft (0.15-0.5 keV) and the hard (0.8-3.0 keV) energy bands.
The total count rate variability fraction\footnote{The count rate variability fraction is defined as (C$_{\rm max}-$ C$_{\rm min}$)/(C$_{\rm max}+$ C$_{\rm min}$) where C$_{\rm min}$ and C$_{\rm max}$ are the minimum and the maximum count rates of the light curve respectively.} is of $\sim$(7$\pm$1)\%. 

No significant orbital variability of \psra's pulsed fraction (or correlation with flux variability) is evident above an upper limit of $\sim$15\% on the full energy band and also in the soft and hard sub-bands.

We noticed that the orbital phases corresponding to the maximum of the \psrb's pulsed flux do not correlate with the phases related to the maximum of the orbital flux (see Fig.\,\ref{fig:psrb_scheme-lc}).

In order to probe possible effects due to intrinsic changes in the binary system (i.e. not related to the observer's line of sight), data were also inspected accounting for the possibility that the emission could rise around periastron epochs due to hypothetical pulsars interactions strengthening. In this case, no appreciable orbital flux variability was detected.

Since radio observations reveal a $\sim$30 s eclipse of \psra\ caused by \psrb's magnetosphere around  \psra\ superior conjunction (phase ph$_{\rm A} \simeq 0.25$), we also searched for possible X-ray variability on such a short time-scale. Given the very high counts statistics available, we calculated that a full $\sim$30 s gap in X-ray \psra\ flux should be in principle detectable at $>$3~$\sigma$ level if present ($\sim$20-30 source counts missed).
No signs of eclipse were found on light curves around \psra's superior conjunction.

\section{Discussion}\label{sec:discussion}

The Double Pulsar is a very complex source of X-ray emission arising from both neutron stars (including magnetospheric and possibly surface pulsed emission) and their peculiar interactions.
\xmm\ provided $\sim$20,000 X-ray counts overall from the system obtained in two dataset ($\sim$590~ks of observing time) separated by a five years gap. During this period \psrb\ disappeared in radio.

Our X-ray timing solution for \psra\ obtained on a 5-years data span represents the most accurate pulsar timing measurement performed by \xmm\ so far, proving phase coherence stability of \pn\ time series on very long timescales.
\citet{mkc+12} provided a systematic check of \xmm\ timing calibration and performances based on long-term X-ray and radio pulsar monitoring. They defined the {\it relative timing accuracy} of an observation as the difference between the period measured with \xmm\ and the period measured at radio wavelengths evaluated at the epoch of the X-ray observations. This difference was normalized to the pulse period measured in radio (see Equation 1 in \citealt{mkc+12}).
The relative timing accuracy of \xmm\ was proved to be better than 10$^{-8}$ on average in their analysis.
Applying the same method to our timing results on \psra\ we found a relative timing accuracy of $\sim$10$^{-12}$, two orders of magnitude smaller than the smallest value found by \citet{mkc+12}.

Detailed study of the radio pulse profile evolution of \psrb\ revealed that the relativistic spin precession was the cause of the radio emission disappearance with respect to our line of sight \citep{pmk+10}, which has been observed in 2008. Although the spin precession rates of \psra\ and \psrb\ are comparable, there is no evidence for secular variation in the radio pulse profile of \psra\ \citep[][2013]{mkp+05,fsk+08}. Results of visual inspection of \psra's 2006 and 2011 X-ray light curves and K-S test (see \S \ref{sec:psra}) confirm this behaviour endorsing that \psra\ does not show clear evidence of precession likely because its spin misalignment from the orbital normal is very small.
Relativistic spin precession of \psra\ should also produce a relative spin phase shift between the 2006 and 2011 light curves of
$\phi \approx \Delta T_{\rm X}/P_{\rm prec} \delta_{\rm A}$ for nearly aligned spin-orbit axes, where $\Delta T_{\rm X}$ is the time between the two observations, $P_{\rm prec}$ is the relativistic spin precession period, and $\delta_{\rm A}$ is the angle between the spin and the orbital angular momentum axes. We noticed that the phase shift minimizing the $\chi^2$ of the ratio between the 2006 and 2011 folded light curves is $\phi \approx 0.008^{+0.001}_{-0.006}$ (1~$\sigma$ errors) corresponding to $\delta_{\rm A} \approx6\degc6^{+1\degc3}_{-5\degc4}$, consistent with the value obtained by \citet{fsk+13}.

From timing analysis we assessed that $\sim$16\% of the source flux is provided by \psrb's pulsed emission, confirming the detection of the pulsar in X-ray even after its radio disappearance in radio in 2008. The peculiar phenomenology of \psrb\ and its weak flux required very careful cross-checks for detection verification.

\psrb\ is clearly detected folding the full 2011 dataset in the 0.15-3 keV energy range, although it shows both pulsed flux and profile variations along the binary orbit possibly including slight spin phase shifts. This very unusual phenomenology is probably due to the
peculiar X-ray brightening process of \psrb\ through energy transfer from \psra\ and provides interesting clues
about the complex pulsar interactions, but prevents the possibility to coherently fold the data over the entire \xmm\ data span.

Discrepancy between the X-ray ($P_{\rm B,X}$ = 2.7734665(8)~s) and radio ($P_{\rm B,radio}$ = 2.77346077007(8)~s) periods of \psrb\ may arise from the time limited validity of radio ephemeris \citep{ksm+06}. Alternatively, that could be ascribed to a physical origin. Changes in rotational parameters could in principle be possible as \psrb\ disappeared in 2008 and we are not able to witness the radio evolution of the pulsar. Glitches and/or timing noise could have occurred in the time between the radio and the second \xmm\ Large Program. An alternative explanation is that the X-ray emission region could be located far from the neutron star surface, affecting the binary orbit timing corrections.
Considering that the light cylinder of \psrb\ is $\sim$$1/3$ of the neutron stars separation, if the emission region were placed in the outer part of the magnetosphere, it is clear that binary timing correction should be strongly affected: a smearing effect on the light curve up to 0.5-1 s (0.2-0.3 in phase) is expected as a function of the orbital phase.
Moreover, the X-ray luminosity of \psrb\ is comparable with its spin-down energy confirming that its brightening is due to \psra\ energy injection. 

The complex modeling of the orbit of this emitting region could in principle allow us to perform binary timing corrections with better accuracy improving the understanding and significance of pulsed signal from \psrb. The timing solution described in this paper could be used to search for peculiar \psrb's pulsed signal in radio archival data arising from that region (even after its radio disappearance according to standard radio ephemeris). Also they will be valuable for studying the eclipse phenomenology.

The analysis of the full \xmm\ dataset provided evidence for the first time of flux orbital variability ($\sim$7\% modulation amplitude).
The observed count excess does not correlate with pulsed flux variations from \psra\ nor from \psrb. The observed feature is likely associated with an additional emission component originating from an interaction layer between the neutron stars. 

\begin{figure*}[t]
\centering
\resizebox{.59\hsize}{!}{\includegraphics[angle=00]{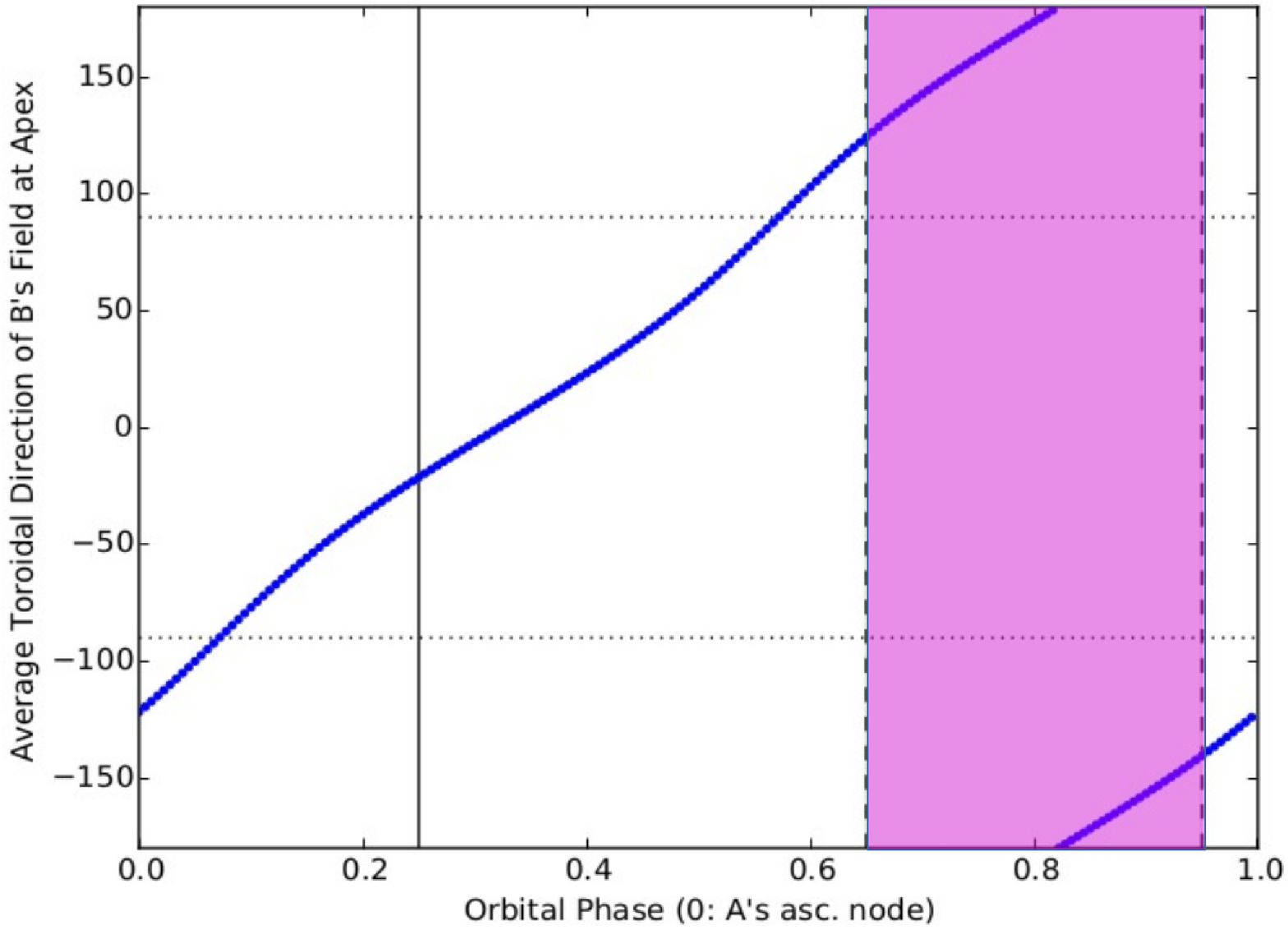}}
\resizebox{.39\hsize}{!}{\includegraphics[angle=00]{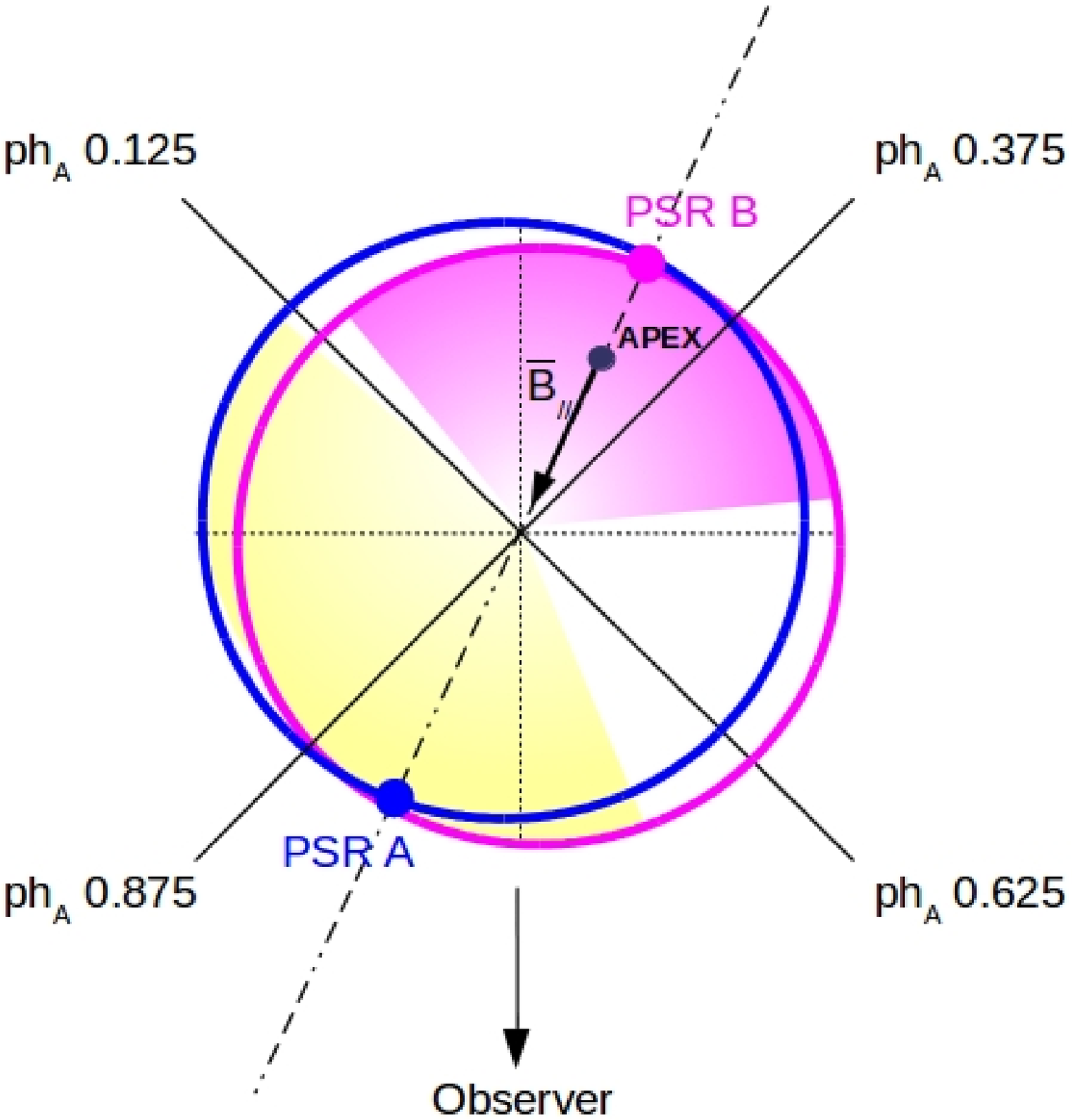}}
\caption{Left panel: component of the \psrb's magnetic field in the orbital plane averaged over one spin period. The pink shaded region indicates where the \psrb's flux reaches its maximum values, while the vertical black line marks the eclipse of \psra. Right panel: scheme of the emission direction of \psrb. The black arrow at the top indicates the orientation of the average toroidal direction of the \psrb's magnetic field, $\overline{B}_{\parallel}$, when \psrb\ is located at the center of its brightest orbital phase interval, indicated by the pink shaded region. While the shaded yellow region represents the range of orbital phases in which the total flux reaches its maximum values.}
\label{fig:Breton}
\end{figure*}

We expect that orbital-dependent persistent X-ray emission comes from the interaction of \psra's wind with \psrb's magnetosphere \citep{liu04}. \psrb\ intersects approximately $ 0.01$ of the \psra\ wind, thus, we expect that the persistent X-ray emission would be $< 0.01 L_A  \sim 6 \times 10^{31}$ erg/s, consistent with observations. 
The X-ray emission from the shock of \psra's wind is expected to be orbital-dependent, since at different orbital phases the shape of the shocked layer of \psra's wind at \psrb's magnetosphere (the magnetosheath) varies, approximately by $\sim 15\%$. In addition, the velocities of the shocked plasma within the magnetosheath are mildly relativistic - this will tend to increase the modulation. 

The pulsed X-ray emission from \psrb\ is the most puzzling. We suggest that in addition to a possible smearing on the light curve of \psrb\ caused by the altitude of the emission region (affecting the binary timing correction), high timing noise is present. This is due both to the orbital variations of the wind-magnetosphere interaction and the resulting turbulence in the magnetosheath. Some of the timing noise may, in fact, be non-random.  \cite{ll14} reproduced many features of the orbital variation of \psrb\ radio emission, like  the orbital variations of intensity,  secular evolution of the radio profile, and  the  subpulse drift  \citep{mkl+04b}, 
by modeling the  distortions of the magnetosphere of \psrb\ by the magnetized wind from  \psra. If the X-ray emission from \psrb\ is produced at high altitudes from the neutron star (see below), we expect variations of X-ray intensity. We suggest that it is related to synchrotron emission of particles from \psra's wind trapped in the magnetosphere of \psrb\ (similar to van Allen belts in the Earth magnetosphere). The model of  radio eclipses of \psra\ \citep{lt05} requires a population of relativistic particles on closed field lines of \psrb\footnote{Plasma density and typical Lorentz factors of plasma particles  cannot be determined  independently, since the requirement of sufficiently high optical depth through \ms\ put a constraint on the product $n \gamma^{-5/3}$, but not the $n$ and $\gamma$ independently}. 

Let us estimate the expected synchrotron luminosity from the \ms\ of \psrb. Synchrotron emission can be produced  by particles only outside the cooling radius, where cyclotron decay time $\tau_{\rm c} \sim m_{\rm e}^3 c^5 / ( B^2 e^4 )$ (with $c$ the light velocity, $B$ the magnetic field, and $e$ and $m_{\rm e}$ the electron charge and mass, respectively) is of the order of the pulsar period,
\be
{ r_{\rm c} } \sim \left( {B_{\rm NS} ^2 e^4 \over m_e ^3 c^5 \Omega_{\rm NS}} \right)^{1/6}  R_{\rm NS} =
2 \times 10^8 {\rm cm}
\ee
where B$_{\rm NS} = 4 \times 10^{11}$ G is the surface \Bf\ of \psrb\  \citep{liu04}, R$_{\rm NS}$ its radius, and $\Omega_{\rm NS}$ the angular velocity.
The synchrotron photon energy emitted at that radius is 
\be
\epsilon = \hbar {e B(r_{\rm c}) \over m_{\rm e} c} \gamma^2= 10^{3} \gamma_{3}^2 {\rm eV}
\ee
where $\gamma_{3}^{}=\gamma/10^3$ is the Lorentz factor of emitting particles.  

The emitting power from the whole \ms\ can then be estimated as
\begin{eqnarray}\nonumber \\
L_X   &\sim& {4 \pi \over 3} r_{\rm c}^3 {e^2 \over c}\Omega_{\rm S} B (r_{\rm c})^3 \gamma^2 n\sim  \lambda B_{\rm NS} R_{\rm NS}^3 \gamma^2 m_{\rm e} c \Omega_{\rm NS}^2        \nonumber \\
   &\sim& 2 \times 10^{31}   \lambda _3 {\rm erg/s} 
\end{eqnarray} \nonumber \\  
where $\lambda_3^{} = \lambda/10^3$ parametrizes the density of emitting electrons with respect to Goldreich and Julian density at $r_{\rm c}$, 
$ n\sim \lambda \Omega_{\rm NS} B(r_{\rm c}) / (2 \pi e c)$.
We estimated $\lambda \sim 10^3$; large multiplicities are  required by the  eclipse modeling \citep{lt05,bkm+12}.

The orbital dependence of the pulsed X-ray emission from \psrb\ may be due to reconnection-induced, orbital phase-dependent penetration of the wind plasma into \psrb's closed field lines. In case of planetary {\ms}s, the plasma entry through the cusp\footnote{The cusp field line separates the field lines that close in the {\it head} and {\it tail} regions of the magnetosphere.} depends on the average angle between the cusp normal and the wind \citep{kal04}; in the Double Pulsar system this average angle depends on the orbital location.
We calculated the average (over one spin period) toroidal direction of \psrb's magnetic field component, $\overline{B}_{\parallel}$, as a function of the orbital phase, where toroidal here refers to the component of the field along the orbital plane of the system. The angle was calculated at the apex of \psrb's magnetosphere, i.e. at its magnetopause boundary along the line connecting it to \psra. A toroidal angle of 0\deg\ indicates that $\overline{B}_{\parallel}$ is pointing away from \psra\ on the line connecting it to \psrb, 180\deg\ is towards \psra, 90\deg\ is parallel to the orbital motion, and $-$90\deg\ is anti-parallel. The result is displayed in the left panel of Fig.\,\ref{fig:Breton}, where the vertical black line marks the eclipse of \psra\ and the pink shaded region is where \psrb\ is found to be brightest in X-rays.
From what is shown in the plot we can infer that \psrb's brightest phase is centered almost exactly around a toroidal angle of 180\deg, where $\overline{B}_{\parallel}$ is pointing toward \psra\ and thus allows particles from the magnetosheath to penetrate more easily into the cusp\footnote{Being toroidal angle 0$^\circ$ disfavored since reconnection (flux transfer events) requires oppositely directed field.}. This outcome strongly supports what we obtained from the observations and it can be visualized as the right panel of Fig.\,\ref{fig:Breton}.

\section{Conclusions}

In order to disentangle the X-ray emission from the two neutron stars and their interactions, the \xmm\ data collected
through two Large Programs on the Double Pulsar require a careful step-by-step approach ranging from relativistic timing analysis to orbital/spin phase-resolved spectral analysis.

In this first paper we presented detailed results on the long-term monitoring of \dbpsr\ allowing us to investigate about
$\sim$80\% of the binary system's total X-ray flux through accurate timing analysis (i.e. from the total count, 60\% and 16\% are \psra\ and \psrb's pulsed emission, respectively, and 7\% is orbital variability).
\xmm\ demonstrated reliable and unprecedent millisecond timing capabilities over a 5-year time span, providing for the first time physical constraints to \psra's light curve stability in X-ray and evidence for orbital flux variability likely associated to the interaction layer between the pulsars. We also provided evidences of X-ray pulsations from \psrb\ even after its radio disappearance. This peculiar phenomenology of PSR B can be ascribed to the particular location of the X-ray emitting region being different from the radio sites, and possibly situated far from the neutron star surface. We demonstrated that the enhanced emission occurs when the average toroidal component of \psrb's magnetic field is pointing toward \psra, which implies that the inflow of new plasma is facilitated.

Relative X-ray flux contributions associated to the above findings provide a first set of inputs for a comprehensive and evolutive
theoretical modeling of the system in X-ray, and offer a useful basis to proceed with spectral analysis and its correlation with timing analysis outcomes (Egron et al. in prep.).

Further monitoring of \dbpsr\ would be suitable for the {\it Suzaku} and {\it NuSTAR} X-ray missions over a few-year interval in order to confirm and study the evolution of \xmm\ findings. For example, {\it NuSTAR} could explore energies $>$6 keV with over an order of magnitude improved sensitivity with respect to \xmm\ looking for pulsars interaction signatures free from strong pulsed soft emission and thermal components.

\acknowledgements
We thank S. Chatterjee for his helpfulness in cross-checking the \psra's multiwavelenght phasing. This work is based on observations obtained using XMM–Newton, an ESA science mission with instruments and contributions directly funded by ESA member states and NASA. Alberto Pellizzoni, Elise Egron, Andrea Possenti, Sara Elisa Motta, Andrea Tiengo and Marta Burgay acknowledge financial support from the Autonomous Region of Sardinia through a research grant under the programme CRP-25399 PO Sardegna FSE 2007-2013, L.R. 7/2007, {\it Promoting scientific research and innovation technology in Sardinia}. The research leading to these results has received funding from the European Union Seventh Framework Programme under grant agreement PIIF-GA-2012-332393.
%

\end{document}